\documentclass[letterpaper,10pt]{article} 

\usepackage{osameet3} 
\usepackage{multicol} 
\usepackage{silence}
\usepackage[font=footnotesize,labelfont=bf]{caption}
\usepackage{amsmath,amssymb}
\usepackage[colorlinks=true,bookmarks=false,citecolor=blue,urlcolor=blue]{hyperref} 

\begin{document}

\title{Digital Interference Mitigation in Space Division Multiplexing Self-Homodyne Coherent Detection}

\author{
	Hanzi Huang\textsuperscript{(1,2)},
	Yetian Huang\textsuperscript{(1,2)},
	Haoshuo Chen\textsuperscript{(2)},
	Qianwu Zhang\textsuperscript{(1)},\\
	Jian Chen\textsuperscript{(1)},
	Nicolas~K.~Fontaine\textsuperscript{(2)},
	Mikael~Mazur\textsuperscript{(2)},
	Roland~Ryf\textsuperscript{(2)},\\
	Junho~Cho\textsuperscript{(2)},
	Yingxiong Song\textsuperscript{(1)}
}

\address{
	\textsuperscript{(1)} Key laboratory of Specialty Fiber Optics and Optical Access Networks, Joint International Research Laboratory of Specialty Fiber Optics and Advanced Communication, Shanghai Institute for Advanced Communication and Data Science, Shanghai University, 200444 Shanghai, China\\
	\textsuperscript{(2)} Nokia Bell Labs, 791 Holmdel Rd., Holmdel, NJ 07733, USA\\
	}
\email{haoshuo.chen@nokia-bell-labs.com}

\copyrightyear{2021}

\begin{abstract}
We propose a digital interference mitigation scheme to reduce the impact of mode coupling in space division multiplexing self-homodyne coherent detection and experimentally verify its effectiveness in 240-Gbps mode-multiplexed transmission over 3-mode multimode fiber.
\end{abstract}

\section{Introduction}
In self-homodyne coherent detection (SHCD), signal and a pilot tone (PT) co-propagate from the transmitter to the receiver where the PT is used as the local oscillator (LO) for coherent detection.
Laser phase noise can be extensively suppressed in SHCD as same light source is applied for the LO and signal carrier.
The coherence property between the PT and signal can be preserved since both experience similar fiber channel variations through co-propagation over the same fiber.
It helps to relax the requirements for narrow linewidth lasers and reduce the complexity and energy cost of the receiver by omitting or simplifying digital-signal processing (DSP) procedures such as frequency offset and carrier phase recovery~\cite{OFC-2020-Th4C.3,OE-25-22-27834}.
To remotely deliver the PT, polarization domain has been explored by multiplexing the PT and signal over the two orthogonal states of polarization (SOP) of single-mode fiber (SMF) \cite{Ruben_SHCD}.
One drawback of polarization division multiplexing is that spectral efficiency (SE) is reduced by 50\% since one polarization must be occupied for transmitting the PT.
To further increase the SE, it was proposed to implement SHCD in space division multiplexing (SDM) systems where one spatial channel delivers the PT while the rest channels are still carrying signals.
SE reduction in SDM-SHCD becomes negligible provided that sufficient spatial channels can be offered \cite{oe-21-2-1561}.
Ideal SHCD uses a high-purity PT at the receiver to avoid interference terms generated after coherent detection.
However, coupling between the spatial channels of the SDM systems can easily cause a polluted LO and degrade the system performance.
It has been shown that inter-core crosstalk limits the performance of SHCD over multi-core fiber (MCF)\cite{oe-21-2-1561} and weakly-coupled multimode fiber (MMF) is needed to reduce inter-mode group crosstalk for SHCD~\cite{OFC-2016-M3C.3}.

In this paper, we propose an interference cancellation-assisted multiple-input multiple-output (MIMO) equalization scheme to mitigate the $1^{\rm st}$ and $2^{\rm nd}$ order interference components introduced by the crosstalk between the spatial channels in SDM-SHCD systems.
The performance of 30-Gbaud QPSK SDM-SHCD transmission over 3-mode MMF is enhanced employing the proposed scheme, where the LP$_{01}$ mode is used for transmitting the PT and all the LP$_{11}$ modes are used as signal channels under a maximum inter-mode group crosstalk of -7~dB.

\section{SDM-SHCD with Digital Interference Mitigation}

\begin{figure}[t!]
  \centering
  \includegraphics[width=0.9\hsize]{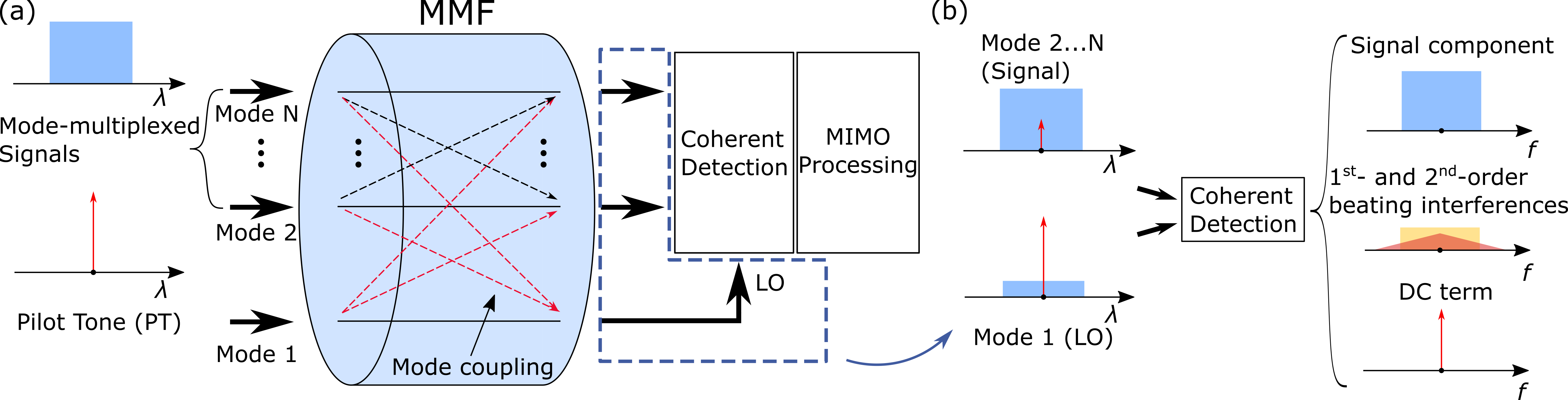}
  \setlength{\abovecaptionskip}{0.1cm}
  \setlength{\belowcaptionskip}{-0.5cm}
\caption{
(a) Schematic of SDM-SHCD systems with mode coupling,
(b) schematic drawing of interference generation after SHCD.
}
\label{fig1}
\end{figure}

Figure~\ref{fig1}(a) shows the schematic of SDM-SHCD.
The red arrows mark the unfavored mode coupling which degrades the purity of the PT due to the mixture between the PT and mode-multiplexed signals.
The first mode group only contains the LP$_{01}$ mode, which makes it a great candidate to deliver the PT with no intra-mode group coupling.  
After MMF transmission, all the modes are demultiplexed into single-mode fibers (SMF) and sent into the signal ports of the coherent receivers except for the LP$_{01}$ mode which is used as the LO.
In conventional SDM transmission using a laser source at the receiver as the LO, mode-multiplexed signals can be recovered employing a linear MIMO equalizer to undo mode coupling \cite{FDE_Equalization,Roland_MIMO}.
Unlike the conventional case, a portion of the PT and signal may both present at the signal and LO port at the receiver due to mode coupling coming from the imperfections from the spatial multiplexer (SMUX) and external perturbations to the MMF.
Using the polluted LO in SHCD, $1^{\rm st}$ and $2^{\rm nd}$ order interference terms together with a strong direct current (DC) term will appear together with the signal components, as illustrated in Fig.~\ref{fig1}(b).
The DC term is from the beating between the PT leaked to the signal and the strong PT at the LO.
The $1^{\rm st}$ order interference term comes from the beating between the PT leaked to the signal and the signal coupled to the LO.
The $2^{\rm nd}$ order interference term can be contributed from the beating between the signals at the signal and LO port.
The $2^{\rm nd}$ order interference term can be better suppressed as a higher PT-to-signal power ratio (PSPR) can be offered.

\begin{figure}[b!]
  \centering
  \includegraphics[width=6.2in]{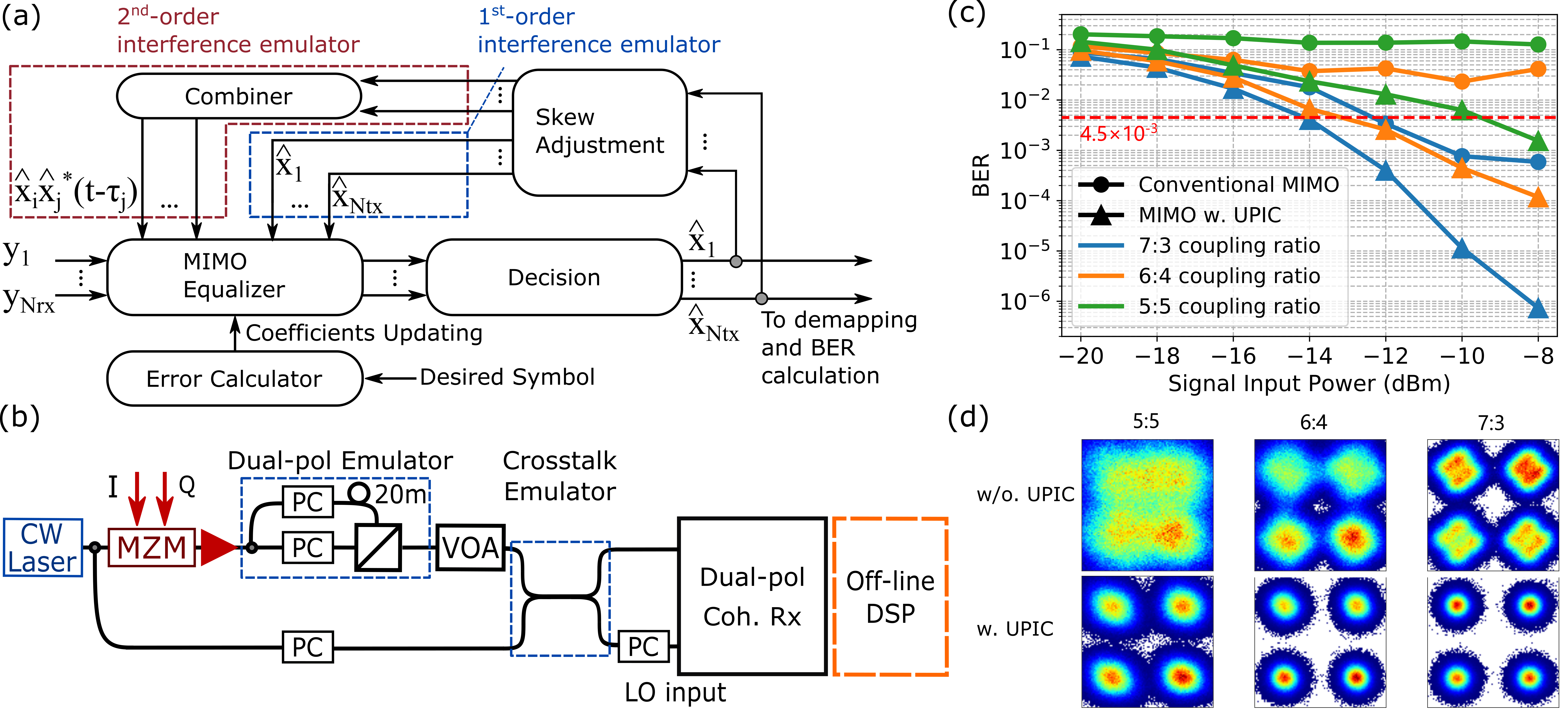}
  \vspace{-0.3cm} 
\caption{
(a) Schematic of the proposed UPIC-assisted MIMO equalization scheme for SDM-SHCD,
(b) experimental setup for verifying the proposed scheme using a 2$\times$2 optical coupler to emulate mode coupling,
(c) measured BER curves versus different input signal power under different coupling ratios,
(d) recovered QPSK constellations of one polarization with and without the aid of UPIC.
}
\label{fig2}
\end{figure}

Figure~\ref{fig2}(a) gives the block diagram of the proposed iterative unreplicated parallel interference cancelling (UPIC)-assisted MIMO equalization scheme.
The key idea of this scheme is to create a cooperative operation between the MIMO equalization and decoding in an iterative manner~\cite{08606153}.
The initial iteration runs a coarse signal detection using a conventional linear MIMO equalization without feedback information.
Starting from the second iteration, the scheme uses feedback symbols to reconstruct the aforementioned interference terms with the help of a skew adjusting and combining module.
The decoded signal becomes more reliable after removing the interference terms with an increased number of iterations.
Soft decoding and error correction code need to be implemented to ensure that a more accurate interference estimation is updated during the iterations. 
In this paper, only one additional iteration is applied assuming the transmitted signal as the correctly decoded feedback symbols to investigate performance upper bound.

\section{Experimental Setup and Results}

To validate the proposed scheme, we first build an experimental setup supporting two spatial channels and use a 2$\times$2 optical coupler to emulate mode coupling, as shown in Fig.~\ref{fig2}(b).
The coupling between the polarization-diversity 10-GBaud Nyquist-shaped QPSK signal and the PT can be varied using an optical coupler with a different coupling ratio.
Two polarization controllers (PC) are added to the PT path.
The PC before the coupler rotates the SOP of the PT by 45 degrees with respect to the SOP of the signal to introduce equal power coupling to each signal polarization and create balanced interference after detection.
The PC at the LO input is to match the SOP of the coherent receiver.
In real applications, photonic integrated polarization tracking devices can be applied~\cite{oe-28-15-21940}.
The LO power at the coupler input port is set to 9~dBm, which makes the PSPR higher enough to negate the $2^{\rm nd}$ order interference.
The measured BER curves as a function of input signal power under different coupling ratios are shown in Fig.~\ref{fig2}(c). 
The QPSK constellations at -8-dBm input signal power are shown in Fig.~\ref{fig2}(d).
Performance is enhanced under all the three coupling conditions, which verifies the effectiveness of the proposed scheme in interference cancellation.


\begin{figure}[t!]
  \centering
  \includegraphics[width=\hsize]{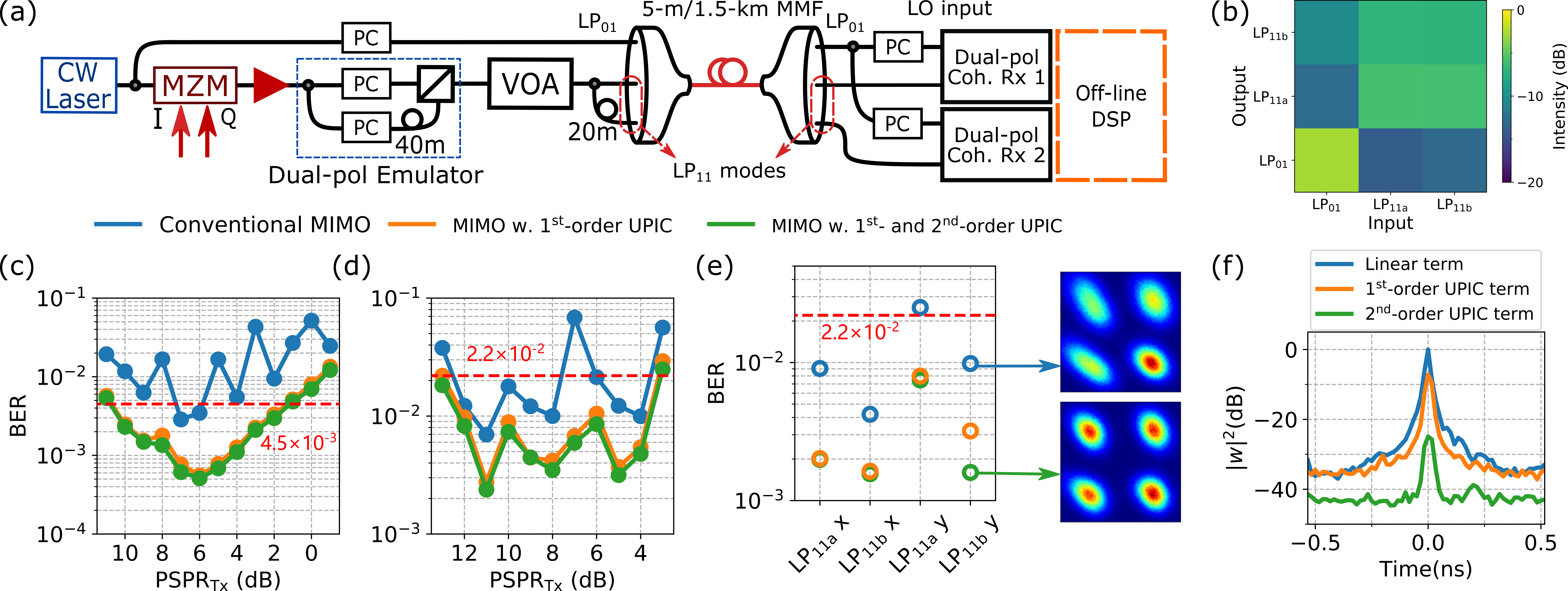}
  \setlength{\abovecaptionskip}{0.0cm}
  \setlength{\belowcaptionskip}{-0.5cm}
\caption{
(a) Setup for SDM-SHCD transmission over 3-mode MMF,
(b) the intensity coupling matrix of the 1.5-km 3-mode MMF link,
(c) measured BER curves employing conventional MIMO and UPIC-assisted MIMO over MMF with a length of 5~m and (d) 1.5~km,
(f) measured BER curves of four modes over 1.5-km MMF at PSPR$_{\rm Tx}$ of 5~dB and QPSK constellations,
(e) tap weights of the UPIC-assisted MIMO equalizer at -1-dBm input signal power.
}
\label{fig3}
\end{figure}

We extend the digital interference mitigation experiment to SDM-SHCD over 3-mode MMF with a length of 5~m (back-to-back, BTB) and 1.5~km using 30-Gbaud Nyquist-shaped QPSK signal as shown in Fig.~\ref{fig3}(a).
A pair of mode group-selective photonic lanterns are used as the SMUXes with a measured mode dependent loss around 5~dB.
Intensity coupling matrix of the 1.5-km 3-mode MMF link is plotted in Fig.~\ref{fig3}(b).
The inter-mode group coupling between the LP$_{01}$ and LP$_{11}$ modes is from -11 to -7~dB, which will dramatically limit the performance of SHCD using conventional MIMO.
Fig.~\ref{fig3}(c) and (d) show the measured BER curves of BTB and 1.5-km 3-mode MMF transmission, which are averaged over the four LP$_{11}$ modes employing three MIMO equalization schemes with and without the aid of UPIC.
Due to dynamically-varying mode coupling, the actual PSPR after transmission cannot be easily obtained.
Therefore, PSPR at the transmitter (PSPR$_{\rm Tx}$) is applied, which is adjusted by attenuating the signal as the power of the PT is fixed.
In the experiment, the PT power is set to 10~dBm for 1.5-km transmission and 9~dBm for BTB case.
For BTB, BER performance can be enhanced by one order of magnitude to be below 7\% forward error correction (FEC) threshold at BER of 4.5$\times10^{-3}$ with the aid of UPIC.
Performance degrades as PSPR$_{\rm Tx}$ decreases below 6~dB, which can be attributed to the increased $2^{\rm nd}$ order interference not fully mitigated using the current algorithm.
For the 1.5-km MMF case, each measurement experiences different mode coupling due to rapid fiber channel changes, which causes different interference conditions and results in varied BER results.
At a low PSPR region and under certain coupling conditions, UPIC-assisted MIMO can effectively mitigate the $2^{\rm nd}$ order interference, see Fig.~\ref{fig3}(e) which shows the BER results of the four LP$_{11}$ modes with a PSPR$_{\rm Tx}$ of 5~dB.
Constellation deformation from the LP$_{\rm 11by}$ mode introduced by strong $2^{\rm nd}$ order interference is extensively mitigated after using the $2^{\rm nd}$ order UPIC.
Fig.~\ref{fig3}(f) gives the time-domain tap weights of the UPIC-assisted MIMO averaged over all the four LP$_{11}$ modes.
The tap weight of the $2^{\rm nd}$ order term is 17~dB smaller than that from the $1^{\rm st}$ order term, which partially explains the marginal performance improvement from the $2^{\rm nd}$ order UPIC.

\section{Conclusion}
We experimentally demonstrated digital interference mitigation in 240-Gbps mode-multiplexed transmission over 3-mode MMF with SHCD under more than -7-dB inter-mode group coupling.
The proposed UPIC-assisted MIMO equalization scheme is scalable to support more spatial channels and beneficial in relaxing the crosstalk constraints in the SDM-SHCD systems which have the potential in high capacity short-reach applications such as intra-datacenter interconnect.

\vspace{5pt}

\scriptsize
\noindent
This work was supported in part by the Science and Technology Commission of Shanghai Municipality (Project No. 20511102400, 20ZR1420900) and 111 project (D20031).

\end{document}